\documentstyle [12pt,epsf]{article}

\input{epsf}
\begin{document}
\begin{titlepage}
\begin{center}

 \vspace{-0.7in}

{\large \bf Stochastic Quantization of Scalar Fields\\in de Sitter Spacetime}\\
 \vspace{.3in}{\large\em T. C. de
 Aguiar\,\footnotemark[1], G. Menezes\,\footnotemark[2] \\
and\\ N.~F.~Svaiter \footnotemark[3]}\\
\vspace{.3in}
Centro Brasileiro de Pesquisas F\'{\i}sicas\,- CBPF,\\
 Rua Dr. Xavier Sigaud 150,\\
 Rio de Janeiro, RJ, 22290-180, Brazil \\

\subsection*{\\Abstract}
\end{center}
\baselineskip .18in

We consider the stochastic quantization method for scalar fields
defined in a curved manifold. The two-point function associated to
a massive self-interacting scalar field is evaluated, up to the
first order level in the coupling constant $\lambda$, for the case
of de Sitter Euclidean metric. Its value for the asymptotic limit
of the Markov parameter $\tau\rightarrow\infty$ is exhibited. We
discuss in detail the covariant stochastic regularization to
render the one-loop two-point function finite in the de Sitter
Euclidean metric.
\\
\vspace{0,3cm}\\
PACS numbers: 03.70+k, 04.62.+v

\footnotetext[1]{e-mail:\,\,deaguiar@cbpf.br}
\footnotetext[2]{e-mail:\,\,gsm@cbpf.br}
\footnotetext[3]{e-mail:\,\,nfuxsvai@cbpf.br}

\end{titlepage}
\newpage\baselineskip .20
in
\section{Introduction}
\quad

The program of stochastic quantization \cite{parisi} and the
stochastic regularization was carried out for generic fields
defined in flat, Euclidean manifolds. For reviews of this program,
see the Refs. \cite{ii} \cite{namiki1} \cite{sakita} \cite{damre}.
In the development of this program some authors applied this
method to linearized Euclidean gravity \cite{gravity1}
\cite{gravity2} and also non-linearized gravity. We may observe,
as it was remarked earlier \cite{menezes3}, that the study of a
situation which lies between these two extremes is missing. A
consistent logical step is to discuss an intermediate situation
between fields in flat spacetime and quantum gravity, i.e., the
semiclassical theory \cite{wald} \cite{ford}.

The stochastic quantization method was used recently to study
self-interacting scalar fields in manifolds which can be
analytically continued to the Euclidean situation, i.e., the
Einstein and the Rindler space \cite{menezes3}. First, these
authors solved a Langevin equation for the mode coefficients of
the field, then they exhibit the two-point function at the
one-loop level. It was shown that it diverges and they used a
covariant stochastic regularization to render it finite. It was
shown that, indeed, the two-point function is regularized. It is
important to remark that this procedure of analytically extend the
real manifold to a complex one in the above situations renders the
action a real quantity, allowing the implementation of the
stochastic quantization in a straightforward way.

Our aim in this article is to discuss the stochastic quantization
of scalar fields defined in a curved manifold, being more
specific, the de Sitter spacetime. For a pedagogical material
discussing the classical geometry of the de Sitter space and the
quantum field theory see, for instance, the Refs. \cite{ford1}
\cite{mottola} \cite{allen} \cite{folacci} \cite{strominger}. Note
that the stochastic quantization is quite different from the other
quantization methods, therefore it can reveal new structural
elements of a theory which so far have gone unnoticed. For
example, a quite important point in a regularization procedure is
that it must preserve all the symmetries of the unregularized
Lagrangian. Many authors have stressed that a priori we can not
expect that a regularization independent proof of the
renormalization of theories in a curved background exists. The
presence of the Markov parameter as an extra dimension lead us to
a regularization scheme, which preserves all the symmetries of the
theory under study. Since the stochastic regularization is not an
action regularization, it may be a way to construct such proof. As
a starting point of this program, for instance, in the
$\lambda\varphi^4$ theory, we should calculate the two-point
function up to the first order level in the coupling constant
$\lambda$ and apply the continuum stochastic regularization. Our
results are to be compared with the usual ones in the literature.

The organization of the paper is the following: in section II we
discuss the stochastic quantization for the
$(\lambda\varphi^{4})_{d}$ scalar theory in a $d$-dimensional
Euclidean manifold. In section III we use the stochastic
quantization and the stochastic regularization to obtain the
two-point Schwinger function in the one-loop approximation in the
Euclidean de Sitter manifold. Conclusions are given in the section
IV. In this paper we use $\hbar=c=k_{B}=G=1$.

\section{Stochastic quantization for the
$(\lambda\varphi^{4})_{d}$ scalar theory: the Euclidean case}

\quad In this section, we give a brief survey for the case of
self-interacting scalar fields, implementing the stochastic
quantization and the continuum stochastic regularization theory up
to the one-loop level. Let us consider a neutral scalar field with
a $(\lambda\varphi^{4})$ self-interaction. The Euclidean action
that usually describes a free scalar field is
\begin{equation}
S_{0}[\varphi\,]=\int d^{d}x\, \left(\frac{1}{2}
(\partial\varphi)^{2}+\frac{1}{2}
m_{0}^{2}\,\varphi^{2}(x)\right), \label{9}
\end{equation}
and the interacting part, defined by the non-Gaussian contribution,
is
\begin{equation}
S_{I}[\varphi]= \int d^{d}x\,\frac{\lambda}{4!} \,\varphi^{4}(x).
\label{10}
\end{equation}

The simplest starting point of the stochastic quantization to
obtain the Euclidean field theory is a Markovian Langevin
equation. Assume a flat Euclidean $d$-dimensional manifold, where
we are choosing periodic boundary conditions for a scalar field
and also a random noise. In other words, they are defined in a
$d$-torus $\Omega\equiv\,T\,^d$. To implement the stochastic
quantization we supplement the scalar field $\varphi(x)$ and the
random noise $\eta(x)$ with an extra coordinate $\tau$, the Markov
parameter, such that $\varphi(x)\rightarrow \varphi(\tau,x)$ and
$\eta(x)\rightarrow \eta(\tau,x)$. Therefore, the fields and the
random noise are defined in a domain: $T\,^{d}\times R\,^{(+)}$.
Let us consider that this dynamical system is out of equilibrium,
being described by the following equation of evolution
\begin{equation}
\frac{\partial}{\partial\tau}\varphi(\tau,x)
=-\frac{\delta\,S_0}{\delta\,\varphi(x)}\bigg|_{\varphi(x)=\varphi(\tau,\,x)}+
\eta(\tau,x), \label{23}
\end{equation}
where $\tau$ is a Markov parameter, $\eta(\tau,x)$ is a random
noise field and $S_{0}[\varphi\,]$ is the usual free Euclidean
action defined in Eq. (\ref{9}). For a free scalar field, the
Langevin equation reads
\begin{equation}
\frac{\partial}{\partial\tau}\varphi(\tau,x)
=-(-\Delta+m^{2}_{0}\,)\varphi(\tau,x)+ \eta(\tau,x),
 \label{24}
\end{equation}
where $\Delta$ is the $d$-dimensional Laplace operator. The Eq.
(\ref{24}) describes a Ornstein-Uhlenbeck process. Assuming a
Gaussian noise distribution we have that the random noise field
satisfies
\begin{equation}
\langle\,\eta(\tau,x)\,\rangle_{\eta}=0 \label{28}
\end{equation}
and
\begin{equation}
\langle\, \eta(\tau,x)\,\eta(\tau',x')\,
\rangle_{\eta}\,=2\delta(\tau-\tau')\,\delta^{d}(x-x'), \label{29}
\end{equation}
where $\langle\,...\rangle_{\eta}$ means stochastic averages. The
above equation for the two-point correlation function defines a
delta-correlated random process. In a generic way, the stochastic
average for any functional of $\varphi$ given by $F[\varphi\,]$ is
defined by
\begin{equation}
\langle\,F[\varphi\,]\,\rangle_{\eta}=
\frac{\int\,[d\eta]F[\varphi\,]\exp\biggl[-\frac{1}{4} \int d^{d}x
\int d\tau\,\eta^{2}(\tau,x)\bigg]}
{\int\,[d\eta]\exp\biggl[-\frac{1}{4} \int d^{d}x \int
d\tau\,\eta^{2}(\tau,x)\bigg]}. \label{36}
\end{equation}
Let us define the retarded Green function for the diffusion
problem that we call $G(\tau-\tau',x-x')$. The retarded Green
function satisfies $G(\tau-\tau',x-x')=0$ if $\tau-\tau'<0$ and
otherwise also
\begin{equation}
\Biggl[\frac{\partial}{\partial\tau}+(-\Delta_{x}+m^{2}_{0}\,)
\Bigg]G(\tau-\tau',x-x')=\delta^{d}(x-x')\delta(\tau-\tau').
\label{25}
\end{equation}
Using the retarded Green function and the initial condition
$\varphi(\tau,x)|_{\tau=0}=0$, the solution for Eq. (\ref{24})
reads
\begin{equation}
\varphi(\tau,x)=\int_{0}^{\tau}d\tau'\int_{\Omega}d^{d}x'\,
G(\tau-\tau',x-x')\eta(\tau',x'). \label{26}
\end{equation}
Let us define the Fourier transforms for the field  and the noise
given by $\varphi(\tau,k)$ and $\eta(\tau,k)$. We have respectively
\begin{equation}
\varphi(\tau,k)=\frac{1}{(2\pi)^\frac{d}{2}} \int\,d^{d}x\,
e^{-ikx}\,\varphi(\tau,x), \label{33}
\end{equation}
and
\begin{equation}
\eta(\tau,k)=\frac{1}{(2\pi)^\frac{d}{2}} \int\,d^{d}x\,
e^{-ikx}\,\eta(\tau,x). \label{34}
\end{equation}
Substituting Eq. (\ref{33}) in Eq. (\ref{9}), the free action for
the scalar field in the $(d+1)$-dimensional space writing in terms
of the Fourier coefficients reads
\begin{equation}
S_{0}[\varphi(k)]\,|_{\varphi(k)=\varphi(\tau,\,k)}=
\frac{1}{2}\int\,d^{d}k\,\varphi(\tau,k)(k^{2}+m_{0}^{2})\varphi(\tau,k).
\label{35}
\end{equation}
Substituting Eq. (\ref{33}) and Eq. (\ref{34}) in Eq. (\ref{24})
we have that each Fourier coefficient satisfies a Langevin
equation given by
\begin{equation}
\frac{\partial}{\partial\tau}\varphi(\tau,k)
=-(k^{2}+m^{2}_{0})\varphi(\tau,k)+ \eta(\tau,k). \label{36}
\end{equation}
In the Langevin equation the particle is subject to a fluctuating
force (representing a stochastic environment), where its average
properties are presumed to be known and also the friction force.
Note that the "friction coefficient" in the Eq. (\ref{36}) is
given by $(k^{2}+m^{2}_{0})$.

The solution for Eq. (\ref{36}) reads
\begin{equation}
\varphi(\tau,k)=\exp\left(-(k^{2}+m_{0}^{2})\tau\right)\varphi(0,k)+
\int_{0}^{\tau}d\tau'\exp\left(-(k^{2}+m_{0}^{2})(\tau-\tau')\right)
\eta(\tau',k). \label{37}
\end{equation}
Using the Eq. (\ref{28}) and Eq. (\ref{29}), we get that the
Fourier coefficients for the random noise satisfy
\begin{equation}
\langle\,\eta(\tau,k)\,\rangle_{\eta}=0 \label{38}
\end{equation}
and
\begin{equation}
\langle\,\eta(\tau,k)\eta(\tau',k')\,\rangle
_{\eta}=2\delta(\tau-\tau')\delta^{d}(k+k'). \label{39}
\end{equation}
It is possible to show that
$\langle\,\varphi(\tau,k)\varphi(\tau',k')\,\rangle_{\eta}|_{\tau=\tau'}\equiv
D(k,k';\tau,\tau')$ is given by:
\begin{equation}
D(k;\tau,\tau)=(2\pi)^d\delta^{d}(k+k')\frac{1}{(k^{2}+m_{0}^{2})}\biggl(1-\exp\left(-2\tau
(k^{2}+m_{0}^{2})\right)\biggr) \label{44}
\end{equation}
where we assume $\tau=\tau'$.

Now let us analyze the stochastic quantization for the
$(\lambda\varphi^{4})_{d}$ self-interaction scalar theory. In this
case the Langevin equation reads
\begin{equation}
\frac{\partial}{\partial\tau}\varphi(\tau,x)
=-(-\Delta+m^{2}_{0}\,)\varphi(\tau,x)-\frac{\lambda}{3!}\varphi^{3}(\tau,x)+
\eta(\tau,x). \label{35}
\end{equation}
The two-point correlation function associated with the random
field is given by the Eq. (\ref{29}), while the other connected
correlation functions vanish, i.e.,
\begin{equation}
\langle\,\eta(\tau_{1},x_{1})\eta(\tau_{2},x_{2})...\eta(\tau_{2k-1},
x_{2k-1})\,\rangle_{\eta}=0, \label{377}
\end{equation}
and also
\begin{equation}
\langle\eta(\tau_{1},x_{1})...\eta(\tau_{2k},x_{2k})\,\rangle_{\eta}=
\sum\,\langle\eta(\tau_{1},x_{1})\eta(\tau_{2},x_{2})\,\rangle_{\eta}
\langle\,\eta(\tau_{k},x_{k})\eta(\tau_{l},x_{l})\,\rangle_{\eta}...,
\label{388}
\end{equation}
where the sum is to be taken over all the different ways in which
the $2k$ labels can be divided into $k$ parts, i.e., into $k$
pairs. Performing Gaussian averages over the white random noise,
it is possible to prove the important formulae
\begin{equation}
\lim_{\tau\rightarrow\infty}
\langle\,\varphi(\tau_{1},x_{1})\varphi(\tau_{2},x_{2})...
\varphi(\tau_{n},x_{n}) \,\rangle_{\eta}= \frac{\int
[d\varphi]\varphi(x_{1})\varphi(x_{2})...\varphi(x_{n})
\,e^{-S(\varphi)}} {\int [d\varphi]\,e^{-S(\varphi)}}, \label{399}
\end{equation}
where $S[\varphi\,]=S_0[\varphi\,]+S_{I}[\varphi\,]$ is the
$d$-dimensional action. This result leads us to consider the
Euclidean path integral measure a stationary distribution of a
stochastic process. Note that the solution of the Langevin
equation needs a given initial condition. As for example
\begin{equation}
\varphi(\tau,x)|_{\tau=0}=\varphi_{0}(x). \label{40}
\end{equation}

Let us use the Langevin equation to perturbatively solve the
interacting field theory. One way to handle the Eq. (\ref{35}) is
with the method of Green's functions. We defined the retarded
Green function for the diffusion problem in the Eq. (\ref{25}).
Let us assume that the coupling constant is a small quantity.
Therefore to solve the Langevin equation in the case of a
interacting theory we use a perturbative series in $\lambda$.
Therefore we can write
\begin{equation}
\varphi(\tau,x)=\varphi^{(0)}(\tau,x)+\lambda\varphi^{(1)}(\tau,x)+
\lambda^{2}\varphi^{(2)}(\tau,x)+... \label{41}
\end{equation}
Substituting the Eq. (\ref{41}) in the Eq. (\ref{35}), and if we
equate terms of equal power in $\lambda$, the resulting equations
are
\begin{equation}
\Biggl[\frac{\partial}{\partial\tau}+(-\Delta_{x}+m^{2}_{0}\,)
\Bigg]\varphi^{(0)}(\tau,x)=\eta(\tau,x), \label{42}
\end{equation}
\begin{equation}
\Biggl[\frac{\partial}{\partial\tau}+(-\Delta_{x}+m^{2}_{0}\,)
\Bigg]\varphi^{(1)}(\tau,x)=-\frac{1}{3!}
\left(\varphi^{(0)}(\tau,x)\right)^{3}, \label{43}
\end{equation}
and so on. Using the retarded Green function and assuming that
$\varphi^{\,(q)}(\tau,x)|_{\tau=0}=0,\,\,\forall\,q$, the solution
to the first equation given by Eq. (\ref{42}) can be written
formally as
\begin{equation}
\varphi^{(0)}(\tau,x)=\int_{0}^{\tau}d\tau'\int_{\Omega}d^{d}x'\,
G(\tau-\tau',x-x')\eta(\tau',x'). \label{sol}
\end{equation}
The second equation given by Eq. (\ref{43}) can also be solved
using the above result. We obtain
\begin{eqnarray}
\varphi^{(1)}(\tau,x)&=&
-\frac{1}{3!}\int_{0}^{\tau}d\tau_{1}\int_{\Omega}d^{d}x_{1}\,
G(\tau-\tau_{1},x-x_{1})\nonumber \\
&&\left(\int_{0}^{\tau_{1}}d\tau'\int_{\Omega}d^{d}x'\,
G(\tau_{1}-\tau',x_{1}-x')\eta(\tau',x') \right)^{3}. \label{44}
\end{eqnarray}
We have seen that we can generate all the tree diagrams with the
noise field contributions. We can also consider the $n$-point
correlation function
$\langle\,\varphi(\tau_{1},x_{1})\varphi(\tau_{2},x_{2})...
\varphi(\tau_{n},x_{n})\,\rangle_{\eta}$. Substituting the above
results in the $n$-point correlation function, and taking the
random averages over the white noise field using the
Wick-decomposition property defined by Eq. (\ref{388}) we generate
the stochastic diagrams. Each of these stochastic diagrams has the
form of a Feynman diagram, apart from the fact that we have to
take into account that we are joining together two white random
noise fields many times. Besides, the rules to obtain the
algebraic values of the stochastic diagrams are similar to the
usual Feynman rules.

As simple examples let us show how to derive the two-point
function in the zeroth order $\langle\,\varphi(\tau_{1},x_{1})
\varphi(\tau_{2},x_{2})\,\rangle^{(0)}_{\eta}$, and also the first
order correction to the scalar two-point-function given by
$\langle\,\varphi(\tau_{1},x_{1})
\varphi(\tau_{2},x_{2})\,\rangle^{(1)}_{\eta}$. Using the Eq.
(\ref{28}), Eq. (\ref{29}) and also Eq. (\ref{26}) we have
\begin{equation}
\langle\,\varphi(\tau_{1},x_{1})
\varphi(\tau_{2},x_{2})\,\rangle^{(0)}_{\eta} =2
\int_{0}^{min(\tau_{1},\tau_{2})}d\tau'\int_{\Omega}d^{d}x'\,
G(\tau_{1}-\tau',x_{1}-x')\,G(\tau_{2}-\tau',x_{2}-x').
 \label{inc1}
\end{equation}
For the first order correction we get:
\begin{eqnarray}
&&\langle\,\varphi(X_{1})
\varphi(X_{2})\,\rangle^{(1)}_{\eta}=\nonumber\\
&& = -\frac{\lambda}{3!}\langle
\int\,dX_{3}\int\,dX_{4}\Biggl(G(X_{1}-X_{4})G(X_{2}-X_{3})+
G(X_{1}-X_{3})G(X_{2}-X_{4})\Bigg)\nonumber\\
&& \eta(X_{3})
\Biggl(\int\,dX_{5}\,G(X_{4}-X_{5})\eta(X_{5})\Bigg)^{3}\rangle_{\eta}.
\label{inc2}
\end{eqnarray}
where, for simplicity, we have introduced a compact notation:
\begin{equation}
\int_{0}^{\tau}d\tau\int_{\Omega}d^{d}x\equiv\int\,dX,
\end{equation}
and also $\varphi(\tau,x)\equiv\varphi(X)$ and finally
$\eta(\tau,x)\equiv\eta(X)$.

The process can be repeated and therefore the stochastic
quantization can be used as an alternative approach to describe
scalar quantum fields. Therefore, the two-point function up to the
first order level in the coupling constant $\lambda$ is given by

\begin{equation}
\langle\,\varphi(\tau_{1},x_{1})
\varphi(\tau_{2},x_{2})\,\rangle^{(1)}_{\eta} = (a) + (b) + (c),
\end{equation}
where $(a)$ is the zero order two-point function and (b) and (c)
are given, respectively, by:
\begin{equation}
(b)=-\frac{\lambda}{2}\,\delta^d(k_1+k_2)\int\,d^dk\int_{0}^{\tau_{1}}\,d\tau\,
G(k_1;\tau_1-\tau)D(k;\tau,\tau)D(k_2;\tau_2,\tau), \label{1}
\end{equation}
\begin{equation}
(c)=
-\frac{\lambda}{2}\,\delta^d(k_1+k_2)\int\,d^dk\int_{0}^{\tau_{2}}\,d\tau\,
G(k_2;\tau_2-\tau)D(k;\tau,\tau)D(k_1;\tau_1,\tau).\\
\end{equation}
These are the contributions in first order. A simple computation
shows that we recover the correct equilibrium result at equal
asymptotic Markov parameters ($\tau_1=\tau_2\rightarrow \infty$):
\begin{equation}
(b)|_{\tau_1=\tau_2\rightarrow
\infty}=-\frac{\lambda}{2}\,\delta^d(k_1+k_2)\frac{1}{(k_2^2+m_0^2)}
\frac{1}{(k_1^2+k_2^2+2m_0^2)} \int\,d^dk\frac{1}{(k^2+m_0^2)}.
\label{700}
\end{equation}

Obtaining the Schwinger functions in the asymptotic limit does not
guarantee that we gain a finite physical theory. The next step is
to implement a suitable regularization scheme. A crucial point to
find a satisfactory regularization scheme is to use one that
preserves the symmetries of the original model. The presence of
the Markov parameter as an extra dimension lead us to a new
regularization scheme, the stochastic regularization method, which
preserves all the symmetries of the theory under study. Therefore,
let us implement a continuum regularization procedure \cite{bh}
\cite{taubes} \cite{c1} \cite{c2} \cite{halp}. We begin with a
regularized Markovian Parisi-Wu Langevin system:
\begin{equation}
\frac{\partial}{\partial\tau}\varphi(\tau,x)
=-\frac{\delta\,S_0}{\delta\,\varphi(x)}\biggl|_{\varphi(x)=\varphi(\tau,\,x)}+
\int d^{d}y\,R_{xy}(\Delta)\, \eta(\tau,y). \label{200}
\end{equation}
Since we are still assuming a Gaussian noise distribution we have
that the random noise field still satisfies the relations given by
Eqs. (\ref{28}) and (\ref{29}). The regulator $R(\Delta)$ that
multiplies the noise is a function of the Laplacian:
\begin{equation}
\Delta_{xy} = \int
d^{d}z\,(\partial_{\mu})_{xz}(\partial_{\mu})_{zy},\label{201}
\end{equation}
where
\begin{equation}
(\partial_{\mu})_{xy} = \partial_{\mu}^x \delta^d(x-y).\label{202}
\end{equation}
In this paper we will be working with a heat kernel regulator with
the form:
\begin{equation}
R(\Delta;\Lambda) = \exp\biggl(\frac{\Delta}{\Lambda^2}\biggr),
\label{203}
\end{equation}
where $\Lambda$ is a parameter introduced to regularize the
theory. The basic restrictions on the form of this heat kernel
regulator are:
\begin{equation}
R(\Delta;\Lambda)|_{\,\Lambda\rightarrow\infty} = 1,
\end{equation}
or
\begin{equation}
R_{xy}(\Delta;\Lambda)|_{\,\Lambda\rightarrow\infty} =
\delta^d(x-y),
\end{equation}
which guarantees that the regularized process given by Eq.
(\ref{200}) reduces to the formal process given by Eq. (\ref{23})
in the formal regulator limit $\Lambda\rightarrow\infty$.

With this modification in the Langevin equation, it is possible to
show that all the contributions to the n-point function at all
orders in the coupling constant $\lambda$ are finite. For
instance, the contribution to the two-point function at the
one-loop level given by Eq. (\ref{1}) is rewritten as:
\begin{equation}
(b)|_{\tau_1=\tau_2\rightarrow
\infty}=-\frac{\lambda}{2}\,\delta^d(k_1+k_2)\frac{R_{k_2}^2}{(k_2^2+m_0^2)}
\frac{1}{(k_1^2+k_2^2+2m_0^2)}
\int\,d^dk\frac{R_{k}^2}{(k^2+m_0^2)}, \label{204}
\end{equation}
where $R_k$ is the Fourier transform of the regulator, i.e.,
\begin{equation}
R_k(\Lambda) = R(\Delta;\Lambda)|_{\Delta = -k^2}.
\end{equation}

Now we use this method to discuss the quantization of scalar
theories with self-interaction in a curved spacetime with event
horizon. Being more specific, we are interested to investigate the
$\lambda\varphi^{4}$ theory in de Sitter Euclidean manifold.

\section{Stochastic quantization for the
$(\lambda\varphi^{4})_{d}$ scalar theory: the de Sitter case}

\quad {\,\,} The aim of this section is to implement the stochastic
quantization and the stochastic regularization for the
self-interacting $\lambda\varphi^{4}$ theory in the one-loop level
in the de Sitter spacetime. Let us consider a $M^{4}$ manifold that
admit a non-vanishing timelike Killing vector field $X$. If one can
always introduce coordinates $t=x^{0},x^{1},x^{2},x^{3}$ locally
such that $X=\frac{\partial}{\partial\,t}$ and the components of the
metric tensor are independent of $t$, $M^{4}$ is stationary. If
further the distribution $X^{\bot}$ of $3$-planes orthogonal to $X$
is integrable, then $M^{4}$ is static. Each integral curve of the
Killing vector vector field $X=\frac{\partial}{\partial\,t}$ is a
world line of an possible observer. Since
$X=\frac{\partial}{\partial\,t}$ generates isometries, the
$3$-planes $X^{\bot}$ are invariant under these isometries. For
static manifold, it is possible to perform a Wick rotation, i.e.,
analytically extend the pseudo-Riemannian  manifold to the
Riemannian domain without problem. Therefore for static spacetime
the implementation of the stochastic quantization is
straightforward.

In the previous section, we have been working in an Euclidean space
$R\,^{d}\times R\,^{(+)}$, where $R\,^{d}$ is the usual Euclidean
space and $R\,^{(+)}$ is the Markov sector. Now let us generalize
this to a more complicated case, i.e., let us work in de Sitter
manifold $M$. In other words, we will consider a classical field
theory defined in a $M\times R\,^{(+)}$ manifold coupled with a heat
reservoir.

In general, we may write the mode decompositions as:
\begin{equation}
\varphi(\tau,x) = \int\,d\tilde{\mu}(k)\varphi_{k}(\tau)u_{k}(x),
\label{216}
\end{equation}
and
\begin{equation}
\eta(\tau,x) = \int\,d\tilde{\mu}(k)\eta_{k}(\tau)u_{k}(x),
\label{217}
\end{equation}
where the measure $\tilde{\mu}(k)$ depends on the metric we are
interested in. For instance, in the flat case, we have that in a
four dimensional space $d\tilde{\mu}(k) = d^{4}k$ and the modes
$u_{k}(x)$ are given by:
\begin{equation}
u_{k}(x) = \frac{1}{(2\pi)^2} e^{ikx}.
\end{equation}

Four-dimensional de Sitter space is most easily represented as the
hyperboloid \cite{davies} \cite{grib}
\begin{equation}
z_0^2 - z_1^2 - z_2^2 - z_3^2 - z_4^2 = -\alpha^2, \label{sit1}
\end{equation}
embedded in five-dimensional Minkowski space with metric
\begin{equation}
ds^2 = dz_0^2 - dz_1^2 - dz_2^2 - dz_3^2 - dz_4^2
\end{equation}
From the form of Eq. (\ref{sit1}), we see that the symmetry group
of de Sitter space is the ten parameter group $SO(1,4)$ of
homogeneous Lorentz transformations in the five-dimensional
embedding space known as the de Sitter group. Just as the
Poincar\'e group plays a central role in the quantization of
fields in Minkowski space, so the de Sitter group of symmetries on
de Sitter space is fundamental to the discussion of its
quantization. The de Sitter space-time $S_{1,3}$ is of constant
curvature. Its Ricci curvature is equal to
$\frac{n(n-1)}{\alpha^{2}}$.

Let us introduce in $S_{1,3}$ the coordinates $x^{\beta} =
{\left(t,\xi^{\,i}\right)}$, where $\beta,\delta,\gamma = 0,1,2,3$
and $i,j=1,2,3$. We are following the Refs. \cite{tag1}
\cite{tag2}. We have:
\begin{equation}
z^0 = \alpha\tan t;\,\,\,\,-\pi/2 < t < \pi/2
\end{equation}
\begin{equation}
z^a = \frac{\alpha}{\cos t} k^a(\xi^1, \xi^2,
\xi^3);\,\,\,\,a,b=1,2,3,4,
\end{equation}
$\xi^1, \xi^2, \xi^3$ being coordinates on the sphere $k_1^2 +
k_2^2 + k_3^2 + k_4^2 = 1$. The infinitely remote ``future"
(``past") corresponds to the value $t = \pi/2$ ($t = -\pi/2$) of
the time-like coordinate $t$. The three-dimensional spheres $z^0 =
const.$ are hypersurfaces of equal time.

If we denote:
\begin{equation}
dk_1^2 + dk_2^2 + dk_3^2 + dk_4^2 = \omega_{ij}(\xi^1, \xi^2,
\xi^3)\,d\xi^i\,d\xi^j,
\end{equation}
where $\omega_{ij} = \frac{\partial k_a}{\partial
\xi_i}\frac{\partial k_a}{\partial \xi_j}$ is the metric tensor of
$S_3$, the interval of the de Sitter space-time is written as
\begin{equation}
ds^2 = \frac{\alpha^2}{\cos^2 t}\biggl(dt^2 - \omega_{ij}(\xi^1,
\xi^2, \xi^3)\,d\xi^i\,d\xi^j\biggr). \label{dsmetric}
\end{equation}
With an usual Wick rotation, we end up in the Euclidean de Sitter
space, written in the coordinates above.

The modes for the conformally coupled scalar field equation for
the Euclidean de Sitter space are given by
\begin{equation}
u_{p s \sigma}^{\pm}(x) = N \sqrt{s+1}  \, \Phi_{s\sigma} \, [k(\xi)]
f_{p}^{\pm}(t) \, \cosh{t}, \label{sit2}
\end{equation}
with $s=0,1,2,...$, $\sigma=1,...,(s+1)^2$. The functions $\Phi_{s\sigma}$ are basic
orthonormal harmonic polynomials in $k$ of degree $s$. They are
labeled by the index $\sigma$. The $f_p^{\pm}(t)$ are expressed
through a hypergeometric function
\begin{equation}
f_{p}^{\pm}(t) =
\frac{1}{(ip)!}\sqrt{\Gamma(ip+\mu)\Gamma(ip-\mu + 1)}\,e^{\pm
ip\,t} F\biggl(\mu,1-\mu;ip + 1;\frac{1\pm \tanh t}{2}\biggr),
\end{equation}
with $\mu = \frac{1}{2}\bigl(1-\sqrt{1-4m^2}\bigr)$ and
$m=m_0\alpha$. The measure for the mode decomposition for Eqs.
(\ref{216}) and (\ref{217}) is given by
\begin{equation}
\int\,d\tilde{\mu}_{p s \sigma} =
\frac{1}{2\pi}\,\int\,dp\,\sum_{s, \, \sigma}. \label{sit3}
\end{equation}
For fields defined in spaces with a general line element given by
\begin{equation}
ds^2 = g_{00}dt^2 + h_{ij}dx^idx^j,
\end{equation}
is possible to prove that the the Parisi-Wu Langevin equation for
scalar fields in Euclidean de Sitter manifold reads
\begin{equation}
\frac{\partial}{\partial\tau}\varphi(\tau,x) =
-\frac{g_{00}}{\sqrt{g}}\,\frac{\delta\,S_0}{\delta\,\varphi(x)}
\bigg|_{\varphi(x)=\varphi(\tau,\,x)}+ \eta(\tau,x), \label{208}
\end{equation}
where $g = \det g_{\mu\nu}$.
The classical Euclidean action $S_{0}$ that appears in the above
equation is given by
\begin{equation}
S_0 = \int\, d^{4}x\,\sqrt{g}\,{\cal{L}},
 \label{209}
\end{equation}
where the Lagrangian density $\cal{L}$ is given by
\begin{equation}
{\cal{L}}=
\frac{1}{2}\,g_{\mu\nu}\,\partial_{\mu}\varphi\,\partial_{\nu}\varphi
+ \frac{1}{2}\,(m_0^2 + \xi R)\varphi^{2}. \label{210}
\end{equation}
Note that we introduce a coupling between the scalar field and the
gravitational field represented by the term $\xi R\varphi^{2}$,
where $\xi$ is a numerical factor and $R$ is the Ricci scalar
curvature. Assuming a conformally coupled field, we have $\xi =
\frac{1}{4}\frac{(n-2)}{(n-1)}$. The random noise field
$\eta(\tau,x)$ obeys the Gaussian statistical law:
\begin{equation}
\langle\,\eta(\tau,x)\,\rangle_{\eta}=0 \label{211}
\end{equation}
and
\begin{equation}
\langle\, \eta(\tau,x)\,\eta(\tau',x')\,
\rangle_{\eta}\,=\,\frac{2}{\sqrt{g(x)}}\,\delta^{4}(x-x')\,\delta(\tau-\tau').
\label{212}
\end{equation}
Substituting Eq. (\ref{209}) and Eq. (\ref{210}) in the Langevin
equation given by Eq. (\ref{208}), we get
\begin{equation}
\frac{\partial}{\partial\tau}\varphi(\tau,x)
=-g_{00}\biggl(-\Delta + m_0^2 +
\frac{2}{\alpha^2}\biggr)\varphi(\tau,x) + \eta(\tau,x).
\label{213}
\end{equation}
Denoting the covariant derivative by $\nabla$, we can define the
four-dimensional Laplace-Beltrami operator $\Delta$ by
\begin{eqnarray}
\Delta &=& g^{-1/2}\partial_{\mu}(g^{1/2}g_{\mu\nu}\partial_{\nu})
\nonumber\\
&=&g_{\mu\nu}\nabla_{\mu}\nabla_{\nu}.
 \label{214}
\end{eqnarray}

To proceed, as in the flat situation, let us introduce the
retarded Green function for the diffusion problem
$G(\tau-\tau',x-x')$, which obeys
\begin{equation}
\Biggl[\frac{\partial}{\partial\tau}+g_{00}\biggl(-\Delta_{x} +
m_0^2 + \frac{2}{\alpha^2}\,\biggr) \Bigg]G(\tau-\tau',x-x')=
\frac{1}{\sqrt{g}}\,\delta^{4}(x-x')\delta(\tau-\tau'),
\label{215}
\end{equation}
if $\tau-\tau'>0$, and $G(\tau-\tau',x,x')=0$ if $\tau-\tau'<0$.

Using the retarded Green function and the initial condition
$\varphi(\tau,x)|_{\tau=0}=0$, a formal solution for Eq.
(\ref{213}) reads
\begin{equation}
\varphi(\tau,x)=\int_{0}^{\tau}d\tau'\int_{\Omega}d^{4}x'\,\sqrt{g(x')}\,
G(\tau-\tau',x-x')\eta(\tau',x'). \label{2666}
\end{equation}

Inserting the modes given by Eq. (\ref{sit2}) in Eq. (\ref{213})
we have that each mode coefficient satisfy the Langevin equation
given by
\begin{equation}
\frac{\partial}{\partial\tau}\varphi_{q}(\tau) =-(q^{2}+
1)\varphi_{q}(\tau)+ \eta_{q}(\tau), \label{226}
\end{equation}
where $q^{2} = p^{2} + \kappa^{2}$ and $\kappa^{2} = s(s+2)$.

The solution for Eq. (\ref{226}), with the initial condition
$\varphi_{q}(\tau)|_{\tau=0}=0$, reads:
\begin{equation}
\varphi_{q}(\tau)= \int_{0}^{\tau}d\tau'G_{q}(\tau,\tau')
\eta_q(\tau'), \label{227}
\end{equation}
where
\begin{equation}
G_{q}(\tau,\tau')= \exp\left(-(q^{2}+ 1)(\tau-\tau')\right)
\theta(\tau-\tau') \label{228}
\end{equation}
is the retarded Green function for the diffusion problem.

Using the relations given by Eq. (\ref{211}) and Eq. (\ref{212}),
we get that the mode coefficients for the random noise satisfies
\begin{equation}
\langle\,\eta_{q}(\tau)\,\rangle_{\eta}=0 \label{229}
\end{equation}
and
\begin{equation}
\langle\,\eta_{q}(\tau)\eta_{q'}(\tau')\,\rangle
_{\eta}=2\delta(\tau-\tau')\delta^4(q,q'), \label{230}
\end{equation}
where $\delta^4(q,q')=\delta(p + p')\delta_{ss'}\delta_{\sigma
\sigma'}$.

The two-point function $D_{q}(\tau,\tau')$ can be calculated in a
similar way as in the Euclidean flat case. We have
\begin{equation}
D_{q}(\tau,\tau')=\frac{1}{(2\pi)}\delta^{4}(q,q')\frac{1}{(q^{2}+
1)} \left(e^{-(q^{2}+ 1)|\tau - \tau'|}- e^{-(q^{2}+ 1)(\tau +
\tau')}\right), \label{231}
\end{equation}
or, in the coordinate representation space:
\begin{eqnarray}
&&D(\tau,\tau';x,x')=
\int\,d\tilde{\mu}(q)\,u^{+}_{q}(x)\,u_{q}^{-}(x')D_{q}(\tau,\tau')
= \nonumber \\
&&\int\,d\tilde{\mu}(q)\,u^{+}_{q}(x)\,u_{q}^{-}(x')\frac{1}{(q^{2}+
1)}\left(e^{-(q^{2}+ 1)|\tau - \tau'|}- e^{-(q^{2}+ 1)(\tau +
\tau')}\right), \label{232}
\end{eqnarray}
where $\int\,d\tilde{\mu}(q)$ is given by Eq. (\ref{sit3}) and,
for simplicity, the index $q$ denotes the set of indices
$ps\sigma$ that label the modes. After a simple calculation
\cite{prud} \cite{grad} it is easy to prove that in the
equilibrium limit $(\tau'=\tau \to \infty)$ we reach the usual
result presented in the literature.

Now, let us apply this method for the case of a self-interacting
theory with an interaction action given by
\begin{equation}
S_{I}[\varphi]= \int d^{4}x\,\sqrt{g(x)}\frac{\lambda}{4!}
\,\varphi^{4}(x). \label{233}
\end{equation}

In the same way as in the Euclidean flat case, we can solve the equation using a perturbative
series in $\lambda$. The two-point function up to the one loop level is given by
\begin{equation}
\langle\,\varphi(\tau_{1},x_{1})
\varphi(\tau_{2},x_{2})\,\rangle^{(1)}_{\eta} = (a) + (b) + (c),
\label{234}
\end{equation}
where $(a)$ is the zero order two-point function given by Eq.
(\ref{231}) and $(b)$ and $(c)$ are given in the momentum space
respectively by
\begin{equation}
(b)=-\frac{\lambda}{2}\,\delta^4(q,k)\int\,d\tilde{\mu}(p)
\, u^+_p \, u^-_p\, \int_{0}^{\tau_{1}}\,d\tau\,
G_{q}(\tau_1-\tau)D_{p}(\tau,\tau)D_{k}(\tau_2,\tau),
\label{235}
\end{equation}
\begin{equation}
(c)= -\frac{\lambda}{2}\,\delta^4(q,k)\int\,d\tilde{\mu}(p) \,
u^+_p \, u^-_p\, \int_{0}^{\tau_{2}}\,d\tau\,
G_{k}(\tau_2-\tau)D_{p}(\tau,\tau)D_{q}(\tau_1,\tau). \label{236}
\end{equation}
These are the contributions in first order. Substituting the
expressions for $G_{q}(\tau-\tau')$ and $D_{q}(\tau,\tau')$
defined in Eqs. (\ref{228}) and (\ref{231}), respectively, one can
easily show that in the asymptotic limit
($\tau_1=\tau_2\rightarrow \infty$) the term $(b)$, for example,
is written as
\begin{equation}
(b)|_{\tau_1=\tau_2\rightarrow
\infty}=-\frac{\lambda}{2}\,\delta^4(q,k) \, \frac{1}{(k^2+1)} \,
\frac{1}{(q^2+k^2+2)} \, I, \label{237}
\end{equation}
\\
where the quantity $I$ is defined as
\begin{equation}
I=\int dp \, \sum_{s=0}^{\infty} \,
\sum_{\sigma=1}^{(s+1)^2} \,\, u_{ps\sigma }^+(x_1)
u_{ps\sigma}^-(x_1) \,\, \frac{1}{p^2+(s+1)^2}.   \label{238}
\end{equation}
Inserting the expression for the modes given by Eq. (\ref{sit2})
and using a summation theorem for harmonic polynomials, the
expression above becomes

\begin{equation}
I=\frac{N^2}{2\pi^2}\cosh^2{t_1} \, \sum_{s=0}^{\infty}
\, (s+1)^3 \int dp \,\,  f^+_p(t_1)\, f^-_p(t_1) \,\, \frac{1}{p^2+(s+1)^2}. \label{239}
\end{equation}
Recalling that the functions $f_p^{\pm}(t)$ are, in fact,
hypergeometric functions, the integral in the expression above can
be performed as long as we choose the appropriate contour in the
$p$ complex plane. These functions are regular in the simple pole
$p_0=-i(s+1)$, and the integration yields

\begin{equation}
I=\frac{N^2}{2\pi}\cosh^2{t_1}\, \sum_{s=0}^\infty
\,(s+1)^2\, f^+_p(t_1)\, f^-_p(t_1)\,\biggl|_{p=-i(s+1)}.\label{240}
\end{equation}

The series in this equation is clearly divergent, so we need a
procedure to regularize it and obtain a finite quantity for the
two-point function. This can be done through the covariant
stochastic regularization procedure. We first introduce a
regularized Langevin equation, which is a generalization of Eq.
(\ref{208}):

\begin{equation}
\frac{\partial}{\partial\tau}\varphi(\tau,x)=-\frac{g_{00}}{\sqrt{g}}
\frac{\delta S_0}{\delta
\varphi(x)}\biggl|_{\varphi(x)=\varphi(\tau,x)}+\int d^4y \,
\sqrt{g} \, R_{xy}(\Delta)\eta(y), \label{241}
\end{equation}
where the regulator is a function of the Laplacian. Using the mode
decomposition given by Eqs. (\ref{216}) and (\ref{217}), the
expression above becomes

\begin{equation}
\frac{\partial}{\partial\tau}\varphi_q(\tau)=-(q^2+1)\varphi_q(\tau)+R_q\eta_q(\tau), \label{242}
\end{equation}
where $R_q=R_{xy}(\Delta)|_{\Delta= -(p^2 + (s+1)^2)}$ and $
R_{xy}(\Delta)= \int\,d\tilde{\mu}(q) \, u^+_{q}(x)u^-_{q}(y) \,
R_{q}$. Then, it is easy to show that the two-point correlation
function for the free field in momentum space is given by

\begin{equation}
D_{q}(\tau,\tau')=\frac{1}{(2\pi)}\delta^{4}(q,q')\frac{R_{q}^2}
{(q^{2}+1)}\left(e^{-(q^{2}+1)|\tau - \tau'|}-
e^{-(q^{2}+1)(\tau + \tau')}\right), \label{243}
\end{equation}
and the first order contribution to the two-point correlation function  is given by

\begin{equation}
(b)_\Lambda|_{\tau_1=\tau_2\rightarrow
\infty}=-\frac{\lambda}{2}\,\delta^4(q,k) \, \frac{R_k^2}{(k^2+1)} \,
\frac{1}{(q^2+k^2+2)} \, I(\Lambda). \label{244}
\end{equation}

The term $I(\Lambda)$ is given by

\begin{equation}
I(\Lambda)=\int dp \, \sum_{s=0}^{\infty} \, \sum_{\sigma=1}^{(s+1)^2} \,\,
u_{ps\sigma }^+(x_1) u_{ps\sigma}^-(x_1) \,\, \frac{R_p^2}{p^2+(s+1)^2}, \label{245}
\end{equation}
which is almost the same expression as the one given by Eq.
(\ref{238}), with the difference that now it is regularized.
Performing the same calculations, one easily obtains
\begin{equation}
I(\Lambda)=\frac{N^2}{2\pi^2}\cosh^2{t_1} \,\int dp\,
\sum_{s=0}^{\infty} \, (s+1)^3 \, \frac{e^{-2(p^2 +
(s+1)^2)/\Lambda^2}}{p^2+(s+1)^2}\, f_{p}^+(t_1) f_{p}^-(t_1).
\label{246}
\end{equation}
The behavior of the exponential function guarantees that the
integral in the above equation is finite. Although we are not able
to present such integral in terms of known functions, to our
purpose it is enough that the integral is finite as we commented.
Therefore, we have obtained the regularized two-point Schwinger
function at the one-loop level for a massive conformally coupled
scalar field in de Sitter space. A similar treatment can be done
to find a regularized four-point Schwinger function also at the
one-loop level.

A quite important point is that this regularization procedure
preserves all the symmetries of the unregularized Lagrangian,
since it is not an action regularization. The next step would be
to isolate the parts that go to infinity in the limit
$\Lambda\rightarrow\infty$ and remove them with a suitable
redefinition of the constants of the theory, i.e., carry out the
renormalization program. A natural question now would be if we can
actually renormalize all the $n$-point functions at all orders at
the coupling constant $\lambda$. Birrel \cite{birrel} has given
arguments that a priori we cannot expect that a regularization
independent proof of the renormalizability of the
$\lambda\varphi^{4}$ theory in a curved background exists. One
attempt of general proof of renormalizability of
$\lambda\varphi^{4}$ theory defined in a spacetime which can be
analytically continued to Euclidean situation was given by Bunch
\cite{bunch}. Using the Epstein-Glaser method, Brunetti and
Fredenhagen \cite{bru} presented a perturbative construction of
this theory on a smooth globally hyperbolic curved spacetime. Our
derivation shows that the stochastic regularization may be an
attempt in a direction of such regularization independent proof,
even though we are still restricted to the same situation studied
by Bunch.

\section{Conclusions and perspectives}

\quad The picture that emerges from these discussions is that the
implementation of the stochastic quantization in curved background
is related to the following fact.  For static manifold, it is
possible to perform a Wick rotation, i.e., analytically extend the
pseudo-Riemannian manifold to the Riemannian domain without
problem. Nevertheless, for non-static curved manifolds we have to
extend the formalism beyond the Euclidean signature, i.e., to
formulate the stochastic quantization in pseudo-Riemannian
manifold, not in the Riemannian space (as in the original
Euclidean space) as was originally formulated. Of course, this
situation is a special case of ordinary Euclidean formulation for
systems with complex actions \cite{menezes1}. It was also shown
that, using a non-Markovian Langevin equation with a colored
random noise, the convergence problem can be solved. It was proved
that it is possible to obtain convergence toward equilibrium even
with an imaginary Chern-Simons coefficient. The same method was
applied to the self-interacting scalar model \cite{menezes2}. We
conclude saying that several alternative methods have been
proposes to deal with interesting physical systems where the
Euclidean action is complex. These methods do not suggest any
systematic analytic treatment to solve the particular difficulties
that arise in each situation.

The possibility of using the stochastic quantization in a generic
curved incomplete manifold is not free of problems. A natural
question that arises in a general situation, or, specifically,
when we are working in Rindler space \cite{rin} \cite{full1}
\cite{full2} \cite{sciama}, is what happens to the noise field
correlation function near an event horizon. It is not difficult to
see that, whenever we have $g = \det g_{\mu\nu} = 0$, this
correlation function diverges, and, therefore, all n-point
correlation functions
$\langle\,\varphi(\tau_{1},x_{1})\varphi(\tau_{2},x_{2})...
\varphi(\tau_{n},x_{n})\,\rangle_{\eta}$ will have meaningless
values, in virtue of the solution of the Langevin equation.
Someone may implement a brick wall-like model \cite{thooft1}
\cite{thooft2} in order to account for these effects; in other
words, one imposes a boundary condition on solutions of the
Langevin equation at a point near the cosmological event horizon
\cite{kim}. On the other hand, in the limit $g = \det
g_{\mu\nu}\rightarrow\infty$, all the correlation functions
vanish. The crucial point of our article was to circumvent the
problem above discussed using a coordinate system where the event
horizon is absent. A natural extension of this paper is to study
the stochastic quantization using a different coordinatization of
the de Sitter space hyperboloid where the event horizon explicitly
appears.

\section{Acknowlegements}

This paper was partially supported by Conselho Nacional de
Desenvolvimento Cientifico e Tecnol{\'o}gico do Brasil (CNPq) and
Funda\c{c}\~ao Carlos Chagas de Amparo \`a Pesquisa do Estado do
Rio de Janeiro (Faperj).

\end{document}